\documentclass[aps,superscriptaddress,showkeys,showpacs,floatfix,preprintnumbers,amsmath,amssymb,12pt]{revtex4}
\usepackage{graphicx} 
\usepackage{epsfig} 
\usepackage{dcolumn}
\usepackage{bm}

\newcommand{\be}{\begin{eqnarray}}
\newcommand{\ee}{\end{eqnarray}}

\begin{document}

\title{Effective pairing interactions with isospin density dependence}

\author{J. Margueron}
\affiliation{ Institut de Physique Nucl\'eaire, Universit\'e
Paris-Sud, IN$_2$P$_3$-CNRS, F-91406 Orsay Cedex, France}
\affiliation{Center for Mathematical Sciences, University of Aizu,
Aizu-Wakamatsu, 965-8580 Fukushima, Japan}
\author{H. Sagawa}
\affiliation{Center for Mathematical Sciences, University of Aizu,
Aizu-Wakamatsu, 965-8580 Fukushima, Japan}
\author{K. Hagino}
\affiliation{Department of Physics, Tohoku University, Sendai, 
980-8578, Japan}

\date{\today}

\begin{abstract}
We perform Hartree-Fock-Bogoliubov (HFB) calculations for 
semi-magic Calcium, Nickel, Tin and Lead isotopes 
and $N$=20, 28, 50 and 82 isotones
using density-dependent pairing interactions recently 
derived from a microscopic nucleon-nucleon interaction. 
These interactions have an isovector component
so that the pairing gaps in symmetric and neutron matter 
are reproduced. 
Our calculations well account for the experimental data 
for the neutron number dependence of binding energy, 
two neutrons separation energy, and odd-even mass staggering
of these isotopes. 
This result suggests that by introducing the isovector term 
in the pairing interaction,
one can construct a global effective pairing interaction which is 
applicable to nuclei in a wide range of the nuclear chart.
It is also shown with the local density approximation (LDA) 
that the pairing field deduced from the pairing gaps in infinite matter 
reproduces qualitatively well the pairing field for finite nuclei 
obtained with the HFB method. 
\end{abstract}



\maketitle


\section{Introduction}

The origin of the pairing correlations in finite nuclei has been under debate 
since the formulation of the BCS theory~\cite{bcs57} and its application to atomic 
nuclei~\cite{boh58,brink-broglia}.
For many-electron systems, the phonon coupling is essential 
in order to get an attractive pairing interaction between electrons. 
In a marked contrast, the nuclear interaction is
already attractive in the $^1$S$_0$ channel even without the phonon
coupling. 
Nevertheless, it has been shown that the phonon coupling in uniform matter
(often referred to as a medium polarization effect) 
as well as in finite nuclei may lead to an important contribution to the nuclear interaction in the
particle-particle channel.
In fact, several many-body methods have been developed
recently for uniform matter in order to include these effects in the calculation
of the pairing gap. These include a group renormalization
method~\cite{sch03}, Monte-Carlo calculations~\cite{fab05,abe07,gez07} and
extensions of the Brueckner theory~\cite{lom99,cao06}.
Most of those calculations have been performed for pure neutron
matter, because of the large interest in the application to neutron star 
physics. 
For instance, the pairing gap is important to understand the cooling of
neutron stars, as it modifies the specific heat as well as some neutrino emission 
processes.
These calculations, except for the one presented in Ref.~\cite{fab05},
predict a reduction of the pairing gap in neutron matter.

It has been known that the pairing correlations play an important role in finite
nuclear systems. 
The relation between finite nuclei and uniform matter, however, 
is not straight forward (see Ref.~\cite{dea03} for a complete review).
In neutron stars, the number of protons is much smaller than that of 
neutrons. No finite nuclei have such extreme proton-to-neutron ratio. 
Also, the density ranges from very low densities up to several times 
nuclear matter saturation density in neutron stars, 
while it is close to the saturation density in finite nuclei. 

Despite these differences, one might view finite nuclei 
as a point in the phase diagram, and extrapolate 
nuclear models to infinite 
matter under the extreme conditions realized in stars. 
Hence, there are mainly two different approaches for a 
calculation of pairing correlations in finite nuclei. 
The first approach is based on phenomenological pairing interactions
whose parameters are determined using some selected data 
and the pairing interaction is usually not uniquely determined 
for the whole nuclear chart (see Refs.~\cite{dob01,dug04} and 
references therein),
while the second approach starts from a bare
nucleon-nucleon interaction and eventually includes the effect of
phonon coupling~\cite{bar99,gio02,bar05}. 
A calculation with the latter approach has recently been carried out, 
based on the nuclear field model. The results of this approach have 
suggested that the medium polarization effects significantly contribute 
to the pairing interaction in finite nuclei
and in fact increase the pairing gap. 

This result is apparently contradict with 
the results in infinite neutron matter,
where the phonon coupling tends to reduce the pairing correlations. 
In order to understand the apparent contradiction, 
an extended Brueckner calculation including the medium polarization effects 
has been performed in Ref.~\cite{cao06} both for symmetric and neutron
matter. 
This calculation has shown that the medium polarization effects act 
differently in neutron matter and in symmetric matter.
That is, the medium polarization effects do not reduce the pairing
gap in symmetric matter, contrary to that in neutron matter.
Instead, in symmetric matter, the neutron pairing gap is 
much enlarged at low density
compared to that of the bare calculation without the polarization effect.
This enhancement takes place especially for 
neutron Fermi momenta $k_{\mathrm{Fn}}<0.7$~fm$^{-1}$.
This provides at least a part of the reason why the medium
polarization effects increase largely the pairing correlation in
finite nuclei but decrease it in neutron matter.

In Ref.~\cite{mar07}, we have proposed effective density-dependent
pairing interactions which reproduce both the neutron-neutron ($nn$) 
scattering length at zero density and the neutron pairing gap in
uniform matter.
In order to simultaneously describe the density dependence of the neutron
pairing gap for both symmetric and neutron matter,
it was necessary to include an isospin dependence in the
effective pairing interaction.
Depending on whether the medium polarization effects on the pairing
gap given in Ref.~\cite{cao06} are taken into account or not, 
we have invented two different density dependences in the pairing
interaction. 
The comparison of predictions of these interactions for finite nuclei with
observed nuclear properties should shed light on 
the links between 
the origin of pairing in finite nuclei and that in
uniform matter. 
This is the main motivation of this work, 
and, in this paper, we apply these interactions to semi-magic nuclei,
such as Ca, Ni, Sn and Pb isotopic chains.
We also investigate isotone chains such as N=20, 28, 50, and 82.

The paper is organized as follows. In Sec.~\ref{sec:ddci} we
briefly remind the main steps for the theoretical HFB approach and 
the procedure we have taken in Ref.~\cite{mar07} to construct the 
density-dependent contact pairing interactions.
Results and predictions for the semi-magic Ca, Ni, Sn and Pb isotopes 
and $N$=20, 28, 50 and 82 isotones
up to the expected drip lines 
are presented in Sec.~\ref{sec:res}. 
In Sec.~\ref{sec:lda}, a local density approximation is
discussed in order to better understand the link with the uniform matter.
Finally, the analysis of the results and the conclusions are given in
Sec.~\ref{sec:conc}. 


\section{Hartree-Fock-Bogoliubov approach with contact density dependent interactions}
\label{sec:ddci}

The self-consistent Hartree-Fock-Bogoliubov (HFB) approach 
in coordinate representation has been presented in detail in Ref.~\cite{doba}.
For the sake of completeness, here we sketch briefly the main part of the model.


\subsection{HFB equations with spherical symmetry}
\label{sec:hfb}

Assuming spherical symmetry and zero range effective nuclear interactions,
the radial HFB equations have the form ($\tau$=n, p):
\begin{widetext}
\begin{equation}
\begin{array}{c}
\left( \begin{array}{cc}
\mathrm{h}_\tau(r) - \lambda_\tau & \Delta_\tau (r) \\
\Delta_\tau (r) & -\mathrm{h}_\tau(r) + \lambda_\tau 
\end{array} \right)
\left( \begin{array}{c} \mathrm{U}_{\tau, i} (r) \\
 \mathrm{V}_{\tau, i} (r) \end{array} \right) = \mathrm{E}_{\tau, i}
\left( \begin{array}{c} \mathrm{U}_{\tau, i} (r) \\
 \mathrm{V}_{\tau, i} (r) \end{array} \right) ~,
\end{array}
\label{eq:hfb}
\end{equation}
\end{widetext}
where $\mathrm{E}_{\tau, i}$ is the quasiparticle energy, $\lambda_\tau$ is the 
chemical potential, $\mathrm{h}_\tau(r)$ is the mean field 
Hamiltonian, and $\Delta_\tau(r)$ is the pairing field. 
The HFB approach consists of solving Eq.~(\ref{eq:hfb}) as a set of
integrodifferential equations with respect to the amplitudes,
$\mathrm{U}_{\tau, i}(r)$ and $\mathrm{V}_{\tau, i}(r)$, as functions of the position coordinate $r$.
In the calculations presented here the mean field Hamiltonian $\mathrm{h}_\tau(r)$
is calculated with the SLy4 Skyrme force~\cite{cha97}, 
and depends on the particle density,
\begin{equation}
\rho_\tau(r) =\frac{1}{4\pi}\sum_{i}(2\mathrm{j}_i+1) \mathrm{V}_{\tau, i}^*(r)
\mathrm{V}_{\tau, i}(r) \; , 
\label{eq:rho}
\end{equation}
as well as on the kinetic and the spin-orbit densities. In Eq.~(\ref{eq:rho}),
the summation is done over the complete space, including
bound and continuum states. 
For the pairing field, we use a density-dependent 
contact force as will be given in Eq.~(\ref{eq:pairing_interaction}) in the
next subsection.
With this force the pairing field is local and is given by:
\begin{equation}
\Delta_\tau(r) = \frac{\mathrm{v}_0 \mathrm{g}_\tau[\rho,\beta]}{2} \tilde{\rho}_\tau (r) \; ,
\end{equation}
where the total density is $\rho=\rho_\mathrm{n}+\rho_\mathrm{p}$, the asymmetry parameter 
is defined as $\beta=(\rho_\mathrm{n}-\rho_\mathrm{p})/\rho$, and the pairing density 
$\tilde{\rho}_\tau(r)$ is
\begin{equation}
\tilde{\rho}_\tau(r)=-\frac{1}{4\pi} \sum_{i} (2\mathrm{j}_i+1) \mathrm{U}_{\tau, i}^* (r) 
\mathrm{V}_{\tau, i} (r) \;.
\label{eq:pdens}
\end{equation}
Because of the nature of the contact interaction,
the pairing density $\tilde{\rho}_\tau$ is divergent, unless a 
cutoff energy is introduced in the sum $i$ of Eq.~(\ref{eq:pdens}).

The continuum states are modelized in this paper as discrete states 
provided by the 
finite-box boundary conditions ($R_\mathrm{box}=25$~fm).
It has been proven that this approximation in the canonical basis
provides an accurate description of the densities and the pairing 
densities~\cite{gra01,sto07}.

\subsection{The density-dependent pairing interactions}
\label{sec:int}

In Ref.~\cite{mar07}, we have taken a contact interaction $\mathrm{v}_{\tau\tau}$
acting on the singlet $^1$S channel,
\be
\langle k|\mathrm{v}_{\tau\tau}|k'\rangle=\frac{1-\mathrm{P}_\sigma}{2}
\mathrm{v}_0 \,\mathrm{g}_\tau[\rho,\beta] 
\,\theta(k,k') \; ,
\label{eq:pairing_interaction}
\ee
where the cutoff function $\theta(k,k')$ is introduced to remove the
ultraviolet divergence in the particle-particle channel.
A simple regularization could be done by introducing a cutoff 
momentum $k_\mathrm{c}$. 
That is, $\theta(k,k')=1$ if $k,k'<k_\mathrm{c}$ and 0 otherwise.
In finite systems, a cutoff energy $e_\mathrm{c}$ is usually introduced
instead of a cutoff momentum $k_\mathrm{c}$.
A detailed discussion on the different prescriptions for the 
cutoff energies in uniform matter are presented in 
Appendix~A of Ref.~\cite{mar07}. 
For a sake of completeness of this paper, we report briefly the
prescription 3 of Ref.~\cite{mar07} which is defined consistently 
with the HFB model.
The cutoff is defined with respect to the quasi-particle energy 
$\sqrt{(\epsilon_\mathrm{n}(k)-\nu_\mathrm{n})^2+\Delta_\mathrm{n}^2}<E_\mathrm{c}$.
This leads to the following definition of the cutoff momenta:
\be
k_\mathrm{c}^\pm = \left[2\mathrm{m}^*
\left(\nu_\mathrm{n}\pm\sqrt{E_\mathrm{c}^2-\Delta_\mathrm{n}^2}\right)\right]^{1/2}/\hbar
\ee
(if $E_\mathrm{c}>\Delta_\mathrm{n}$).
If $k_\mathrm{c}^-$ becomes
imaginary for very small $\nu_\mathrm{n}$,
we set $k_\mathrm{c}^-=0$.
The parameters of the pairing interactions have been obtained
within this prescription.

In Eq.~(\ref{eq:pairing_interaction}), the interaction strength $\mathrm{v}_0$ is 
determined from the low-energy neutron-neutron scattering
phase-shift~\cite{ber91,esb97,gar99,mar07},
that fixes the relation between $\mathrm{v}_0$ and the cutoff energy $e_\mathrm{c}$,
while the density-dependent term $\mathrm{g}_\tau[\rho,\beta]$ is deduced from 
the realistic nucleon-nucleon interaction calculations
of the pairing gaps in symmetric and neutron matter.
The isospin symmetry breaking of the bare nucleon nucleon 
interaction is neglected.
The density-dependent term accounts for the medium 
effects and satisfies the 
boundary condition $\mathrm{g}_\tau\rightarrow 1$ for $\rho\rightarrow 0$.
In Ref.~\cite{mar07}, we have introduced an isovector
dependence in the density-dependent term $\mathrm{g}_\tau[\rho,\beta]$
as $\mathrm{g}_\tau=\mathrm{g}_\tau^1+\mathrm{g}^2$.
In the neutron pairing channel, the term $\mathrm{g}_\mathrm{n}^1$ is given as
\be
\mathrm{g}_\mathrm{n}^1[\rho,\beta] =  1
-\mathrm{f}_\mathrm{s}(\beta)\eta_\mathrm{s}
\left(\frac{\rho}{\rho_0}\right)^{\alpha_\mathrm{s}}
-\mathrm{f}_\mathrm{n}(\beta)\eta_\mathrm{n}
\left(\frac{\rho}{\rho_0}\right)^{\alpha_\mathrm{n}} \;, 
\label{eq:g1n}
\ee
where $\rho_0$=0.16~fm$^{-3}$ is the saturation density of symmetric nuclear
matter, and the term $\mathrm{g}^2$ is added only to the interaction
IS+IV~Induced and is given by
\be
\mathrm{g}^2[\rho]=\eta_{2}\left[\left(1+e^{\frac{k_\mathrm{F}-1.15}{0.05}}\right)^{-1}-
\left(1+e^{\frac{k_\mathrm{F}-0.1}{0.08}}\right)^{-1}\right]
\; .
\label{eq:g2}
\ee
Notice the slight modifications of the parameters in Eq.~\ref{eq:g2}
compare to~\cite{mar07}.
In the proton pairing channel, the isospin symmetry 
of the matrix element (\ref{eq:pairing_interaction})
gives rise to the relation
\be
\mathrm{g}_\mathrm{p}^1[\rho,\beta]=\mathrm{g}_\mathrm{n}^1[\rho,-\beta] \; .
\label{eq:g1p}
\ee
The goal of the functional form in Eqs.~(\ref{eq:g1n}) and
(\ref{eq:g2}) is to reproduce the theoretical calculation of the
pairing gap in both symmetric and neutron matter and also to be used
for prediction of the pairing gap in asymmetric matter.
In finite nuclei, the densities $\rho_\mathrm{n}$ and $\rho_\mathrm{p}$ acquire an
explicit dependence on the coordinate $r$, which defines the density
$\rho(r)$ and the asymmetry parameter $\beta(r)$.
In Eq.~(\ref{eq:g1n}), the interpolation functions $\mathrm{f}_\mathrm{s}(\beta)$ and
$\mathrm{f}_\mathrm{n}(\beta)$ should satisfy the following conditions, 
$\mathrm{f}_\mathrm{s}(0)=\mathrm{f}_\mathrm{n}(1)=1$ and 
$\mathrm{f}_\mathrm{s}(1)=\mathrm{f}_\mathrm{n}(0)=0$. 
It should however be noticed that the interpolation functions $\mathrm{f}_\mathrm{s}(\beta)$
and $\mathrm{f}_\mathrm{n}(\beta)$ cannot be deduced from the adjustment of the pairing gap
in symmetric and neutron matter.
In this paper, we choose 
$\mathrm{f}_\mathrm{s}(\beta)=1-\mathrm{f}_\mathrm{n}(\beta)$ and 
$\mathrm{f}_\mathrm{n}(\beta)=\beta$.

\begin{table}[tb]
\begin{center}
\setlength{\tabcolsep}{.06in}
\renewcommand{\arraystretch}{1.5}
\begin{tabular}{ccccccc}
\toprule 
interaction& $E_\mathrm{c}$ & $\eta_\mathrm{s}$ & $\alpha_\mathrm{s}$ & $\eta_\mathrm{n}$ & $\alpha_\mathrm{n}$ & $\eta_2$ \\
\colrule
\hbox{IS+IV~Bare} & 40~MeV & 0.664 & 0.522 & 1.01  & 0.525 & 0.0 \\
\hbox{IS+IV~Induced} & 40~MeV & 1.80 & 0.27 & 1.61 & 0.122 & 0.8 \\
\hbox{IS~Bare} & 40~MeV & 0.664 & 0.522 & 0.664  & 0.522 & 0.0 \\
\botrule
\end{tabular}
\end{center}
\caption{Parameters for the density-dependent functions, $\mathrm{g}_1$ and $\mathrm{g}_2$
defined in Eqs.~(\ref{eq:g1n}) and (\ref{eq:g2}).
These parameters are obtained from the fit to the pairing gaps in 
symmetric and neutron matter obtained by the microscopic nucleon-nucleon 
interaction. See the text for details.}
\label{tab2}
\end{table}%

We adjust the parameters of the contact pairing interaction so that
the position and the absolute value of the maxima of the pairing gaps 
of the nucleon-nucleon interaction in 
symmetric and neutron matter are reproduced.
For the bare pairing gap, the maximum is located at $k_\mathrm{Fn}=0.87$~fm$^{-1}$
with $\Delta_\mathrm{n}$=3.1~MeV for both symmetric and neutron matter, 
while for the screened pairing gap, the maximum
is at $k_\mathrm{Fn}=0.60$~fm$^{-1}$ with $\Delta_\mathrm{n}$=2.70~MeV for symmetric matter and
$k_\mathrm{Fn}=0.83$~fm$^{-1}$ and $\Delta_\mathrm{n}$=1.76~MeV for neutron matter. 
We call the interaction fitted to the bare pairing gap the IS+IV~Bare
interaction, while that to the screened gap the IS+IV~Induced interaction. 
In order to estimate the importance of the isovector term of the interaction,
we have also parameterized a pure isoscalar interaction, IS~Bare, so as to
reproduce the bare pairing gap in symmetric matter.
The obtained parameters are given in Table~\ref{tab2}.
The best agreement with the results of the microscopic nucleon-nucleon
interaction in Ref.~\cite{cao06} is obtained
with a cutoff energy $E_\mathrm{c}=40$~MeV~\cite{mar07}. 


\section{Results for finite nuclei}
\label{sec:res}

\begin{figure}[b]
\begin{center}
\includegraphics[scale=0.33]{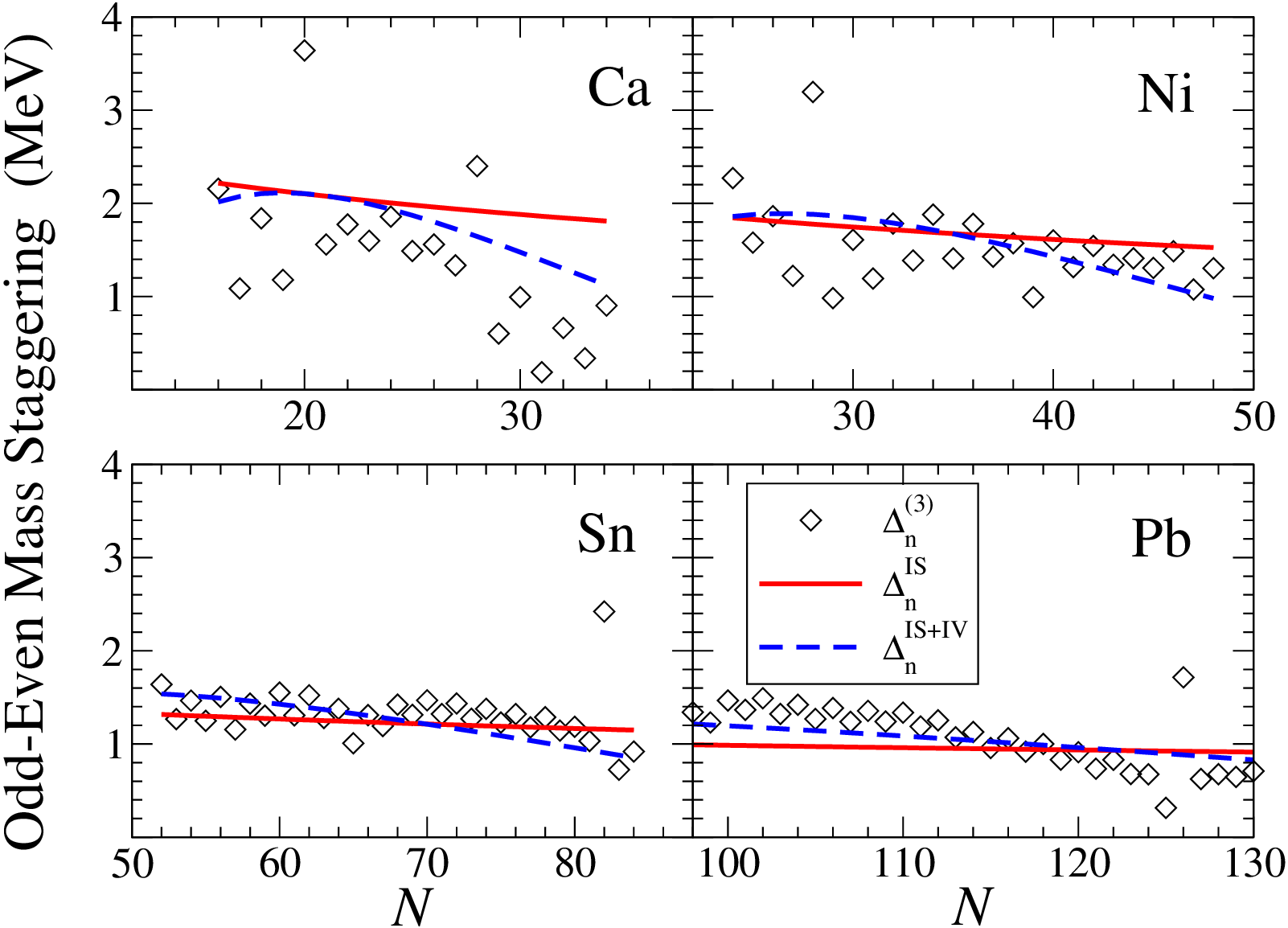}
\end{center}
\caption{(Color online) 
The experimental odd-even mass staggering $\Delta_\mathrm{n}^{(3)}$ given
by Eq.~(\ref{eq:oes}) for the semi-magic Ca, Ni, Sn and Pb isotopes. 
It is compared with the phenomenological fits 
$\Delta_\mathrm{n}^\mathrm{IS}=13.3/A^{1/2}$~MeV and 
$\Delta_\mathrm{n}^\mathrm{IS+IV}=[7.2-44(1-2Z/A)^2]/A^{1/3}$~MeV
proposed in Ref.~\cite{vog84}.}
\label{fig01}
\end{figure}

It is a rather difficult task to extract the pairing gaps from 
the experimental data to compare with the theoretical results
(see for instance Ref.~\cite{dug01} and references therein).
In the following, we thus compare the predictions of the pairing
interactions with different experimental data~\cite{ame2003}: 
the masses per particle $\mathrm{B}(N,Z)/A$, two neutrons separation 
energies defined as $\mathrm{S}_\mathrm{2n}=\mathrm{B}(N,Z)-\mathrm{B}(N-2,Z)$, and the 
odd-even mass staggering (OES) defined as 
\be
\Delta^{(3)}(N,Z)&\equiv&-\frac{\pi_N}{2}\Big[\mathrm{B}(N-1,Z)-2\mathrm{B}(N,Z)
\nonumber \\
&&+\mathrm{B}(N+1,Z)\Big] \; ,
\label{eq:oes}
\ee
where $\pi_N=(-)^N$ is the number parity.
For even nuclei, the OES is known to be sensible not only to the
pairing gap, but also to shell effects and
deformations~\cite{sat98,dug01}.
Therefore, the comparison of a theoretical pairing gap with OES should 
be done with caution.
At a shell closure, the OES~(\ref{eq:oes}) does not go to zero as expected,
but it increases 
substantially (see  Fig.~\ref{fig01}). 
This large gap is an artifact due to the shell effect, which is totally
independent of the pairing gap itself.
In the following, we shall thus remove 
all the nuclei at the shell closures from the comparison to
experimental OES.

The effects of the isospin asymmetry on the 
pairing gap has been suggested for a long time.
In Ref.~\cite{vog84}, the mass number dependence of the pairing gap 
has been extracted from the experimental OES
for nuclei outside the shell closures within the range $50<Z<82$ and
$82<N<126$. 
Two phenomenological fits have been suggested. 
The first one, which we call isoscalar, is only dependent on the mass number $A$,
and reads $\Delta_\mathrm{n}^\mathrm{IS}=13.3/A^{1/2}$~MeV. 
On the other hand, the second one, which we call isovector, 
has a quadratic dependence on the neutron-proton
asymmetry, and is expressed as
$\Delta_\mathrm{n}^\mathrm{IS+IV}=[7.2-44(1-2Z/A)^2]/A^{1/3}$~MeV.
We represent in Fig.~\ref{fig01} the experimental OES
$\Delta_\mathrm{n}^{(3)}$ in Eq.~(\ref{eq:oes}) together with the 
phenomenological fits, $\Delta_\mathrm{n}^\mathrm{IS}$ and
$\Delta_\mathrm{n}^\mathrm{IS+IV}$.
From comparisons between the fits and the experimental OES  
in Fig.~\ref{fig01}, it is difficult to extract the quadratic dependence 
of the pairing gap. 
Namely, the fits $\Delta_\mathrm{n}^\mathrm{IS}$ and 
$\Delta_\mathrm{n}^\mathrm{IS+IV}$ reproduce the experimental OES equally well
in general, despite an appreciable difference in the predictions in very
heavy isotopes. 
The fitting functions are supposed to describe the smooth behavior of the pairing
gaps with $A$ and $Z$, but are not able to describe the fine structure
of the pairing gap in a single nuclei. 
For instance, the drop of the pairing gap at a shell closure is
totally absent.

\begin{figure*}[tb]
\begin{center}
\includegraphics[scale=0.5]{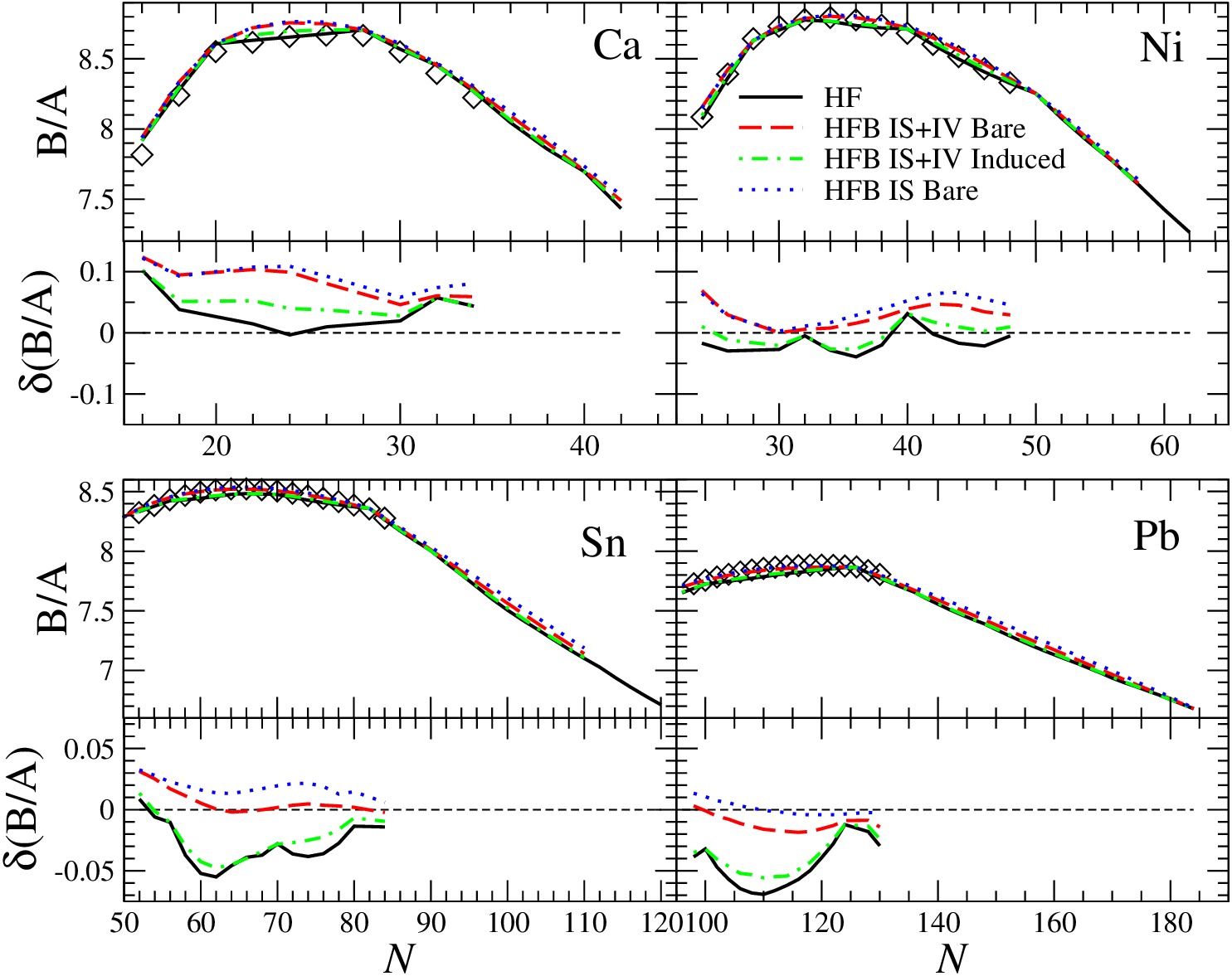}
\end{center}
\caption{(Color online) Comparison of the HFB calculations with the
experimental binding energies, $\mathrm{B}/A$. 
The solid line shows the results without pairing interaction (HF), 
while the dotted, short dashed, and long dashed lines are obtained
with the pairing interactions IS+IV~Bare, IS+IV~Induced and IS~Bare,
respectively. 
For each isotopic chain, we also plot the difference
$\delta(\mathrm{B}/A)=\mathrm{B}(\mathrm{th.})/A-\mathrm{B}(\mathrm{exp.})/A$ between the
theoretical and the exprimental values for the binding energy.
All units are given in MeV.
See the text for more details.}
\label{fig10}
\end{figure*}

The experimental binding energies per particle $\mathrm{B}/A$ is compared
with our results for the two pairing interactions, IS+IV~Bare and 
IS+IV~Induced, in Fig.~\ref{fig10}.
The results can be 
classified into two groups: the first group is the one of the light
isotopes (Ca and Ni) for which the HF calculation is already close 
to the experimental masses, while the second group is the
one of the heavier isotopes (Sn and Pb) for which HF calculations
underestimate the binding energies (see the solid line). 
When the pairing is switched on, the interaction IS+IV~Induced reproduces the
experimental masses in the first group of isotopes within the same
accuracy of HF, while the interaction IS+IV~Bare overestimates the masses
for this group. 
In contrast, the second group of isotopes behaves in an opposite way: 
the pairing interaction IS+IV~Bare leads to masses which are much 
closer to the experimental ones as compared to the IS+IV~Induced 
interaction, or to the HF calculation. 
The interaction IS~Bare will be discussed latter on.
When one compares the difference between the theoretical and
the experimental binding energies $\delta(\mathrm{B}/A)$, it is observed that
the pairing interaction IS+IV~Bare flattens this difference as a function of
neutron number, even for the first group of light isotopes. 
These results suggest that the different behavior between the first 
and the second groups, rather than a pairing effect,
originates from an effect of the mean field Skyrme
interaction (SLy4), which has been parameterized so as to reproduce
better the binding energies of intermediate and heavy 
nuclei rather than those of the light ones.

\begin{figure*}[tb]
\begin{center}
\includegraphics[scale=0.55]{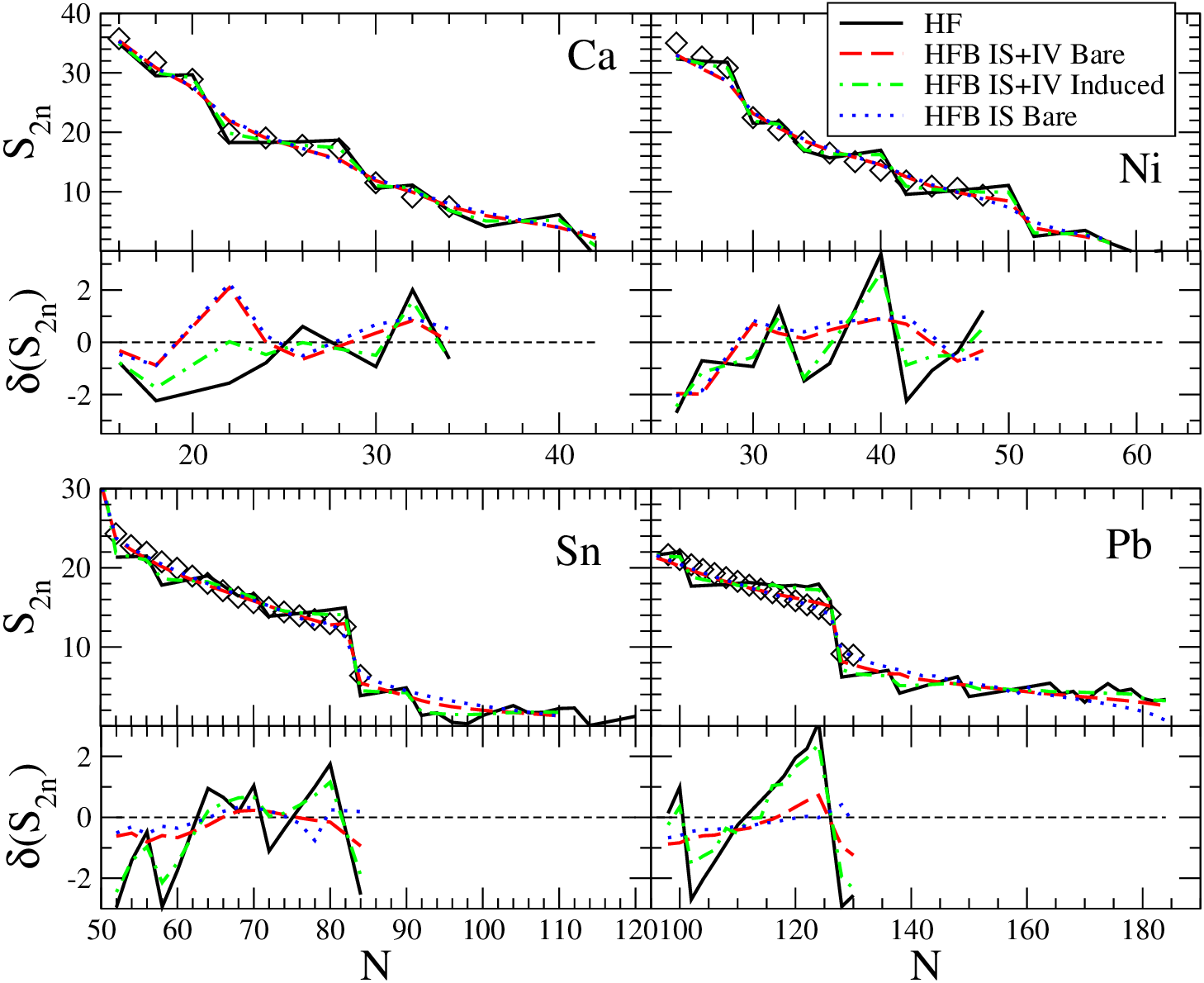}
\end{center}
\caption{(Color online) Comparison between HFB and experiments
for the two-neutron separation energies $\mathrm{S}_\mathrm{2n}$.
The value $\delta(\mathrm{S}_\mathrm{2n})$ is defined as 
$\delta(\mathrm{S}_\mathrm{2n})=\mathrm{S}_\mathrm{2n}(\mathrm{th.})-\mathrm{S}_\mathrm{2n}(\mathrm{exp.})$.
See the caption of Fig.~\ref{fig10} and the text for details.}
\label{fig11}
\end{figure*}

The effect of the pairing correlations can be clearly seen 
in the two neutrons separation
energy $\mathrm{S}_\mathrm{2n}$, which is sensible to the relative difference in
binding energies, and somehow reduces the effect of the mean field
interaction. The results of HFB calculations for $\mathrm{S}_\mathrm{2n}$ are shown in 
Fig.~\ref{fig11} in comparison with experimental data. 
We now see that the pairing interaction IS+IV~Bare works better than 
HF or the IS+IV~Induced interaction for all the four selected isotopic chains.
The dependence of $\mathrm{S}_\mathrm{2n}$ on the neutron number $N$ is much improved
using the interaction IS+IV~Bare than IS+IV~Induced, even for the group of light
isotopes. 
It should be reminded that no tuning for each isotopes has been done for any of
these pairing interactions.

\begin{figure*}[tb]
\begin{center}
\includegraphics[scale=0.55]{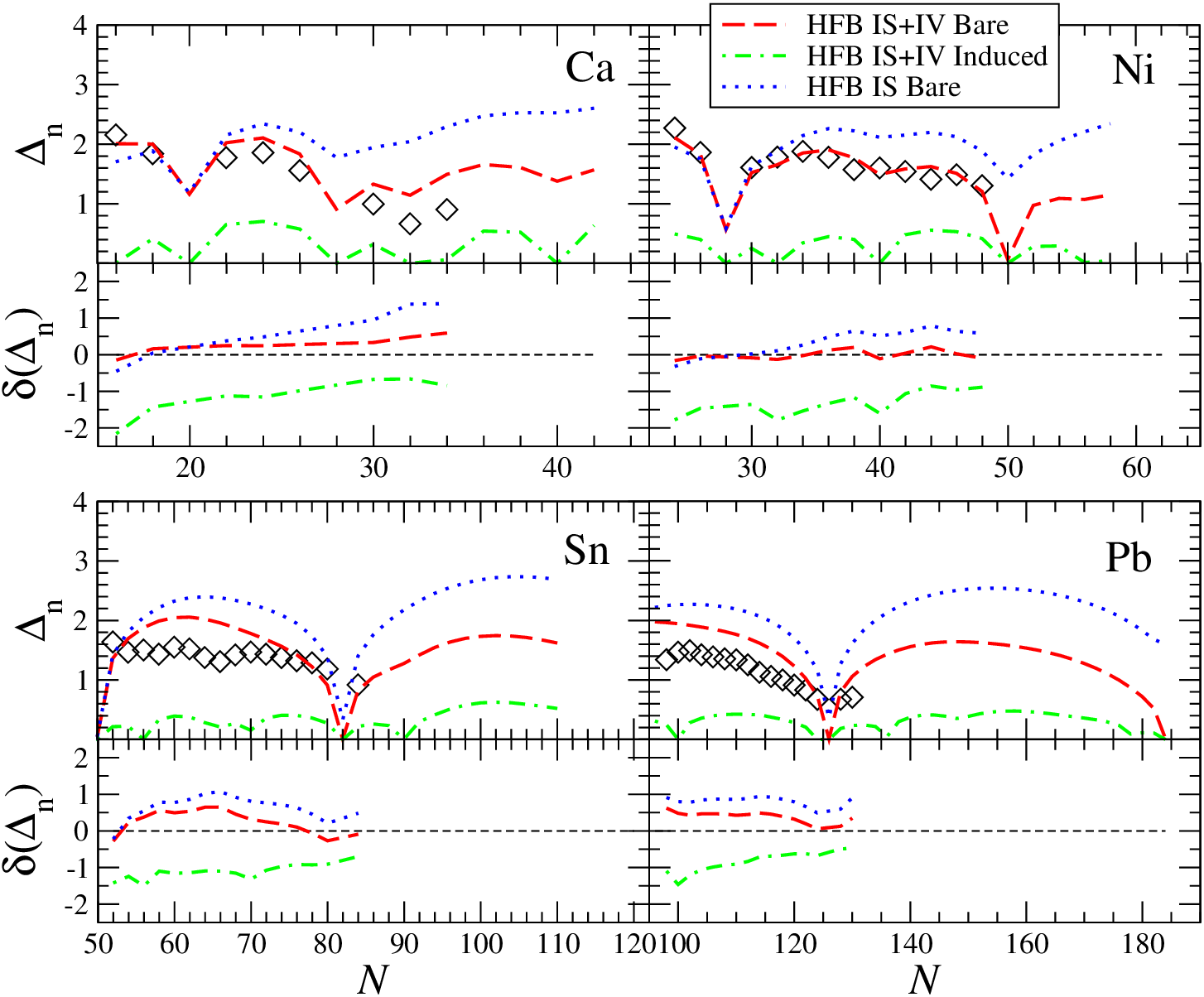}
\end{center}
\caption{(Color online) Comparison of the HFB pairing gaps $\Delta_\mathrm{n}$ calculated with
Eq.~(\ref{eq:delta_mean}) with the OES given by $\Delta^{(3)}$.
The value $\delta(\Delta_\mathrm{n})$ is defined as 
$\delta(\Delta_\mathrm{n})=\Delta_\mathrm{n}(\mathrm{th.})-\Delta_\mathrm{n}(\mathrm{exp.})$.
See the caption of Fig.~\ref{fig10} and the text for details.}
\label{fig12}
\end{figure*}

Let us next compare the experimental OES with the
mean pairing gap calculated from the pairing field $\Delta_\mathrm{n}(r)$ as,
\be
\Delta_\mathrm{n}\equiv\frac{1}{N}\int d^3r \rho_\mathrm{n}(r) \Delta_\mathrm{n}(r) \; , 
\label{eq:delta_mean}
\ee
where $N=\int d^3r\rho_\mathrm{n}(r)$ is the number of neutrons. 
In the next section, we will discuss also another formula for the 
mean pairing gap. The results are shown in Fig.~\ref{fig12}. 
We remind the reader that this comparison should be taken with caution and we
have removed from the comparison the OES calculated at the neutron shell
closure. 
It is observed that the pairing gaps obtained with the interaction 
IS+IV~Induced are systematically too small along the
isotopic chains. This is the reason why
the results with the interaction IS+IV~Induced are 
close to the HF calculations in Figs.~\ref{fig10} and \ref{fig11}.
Contrary, the results with the interaction IS+IV~Bare are in good 
agreement with the experimental OES, including the isotopic trend 
for all the four isotopic chain.

\begin{figure*}[tb]
\begin{center}
\includegraphics[scale=0.55]{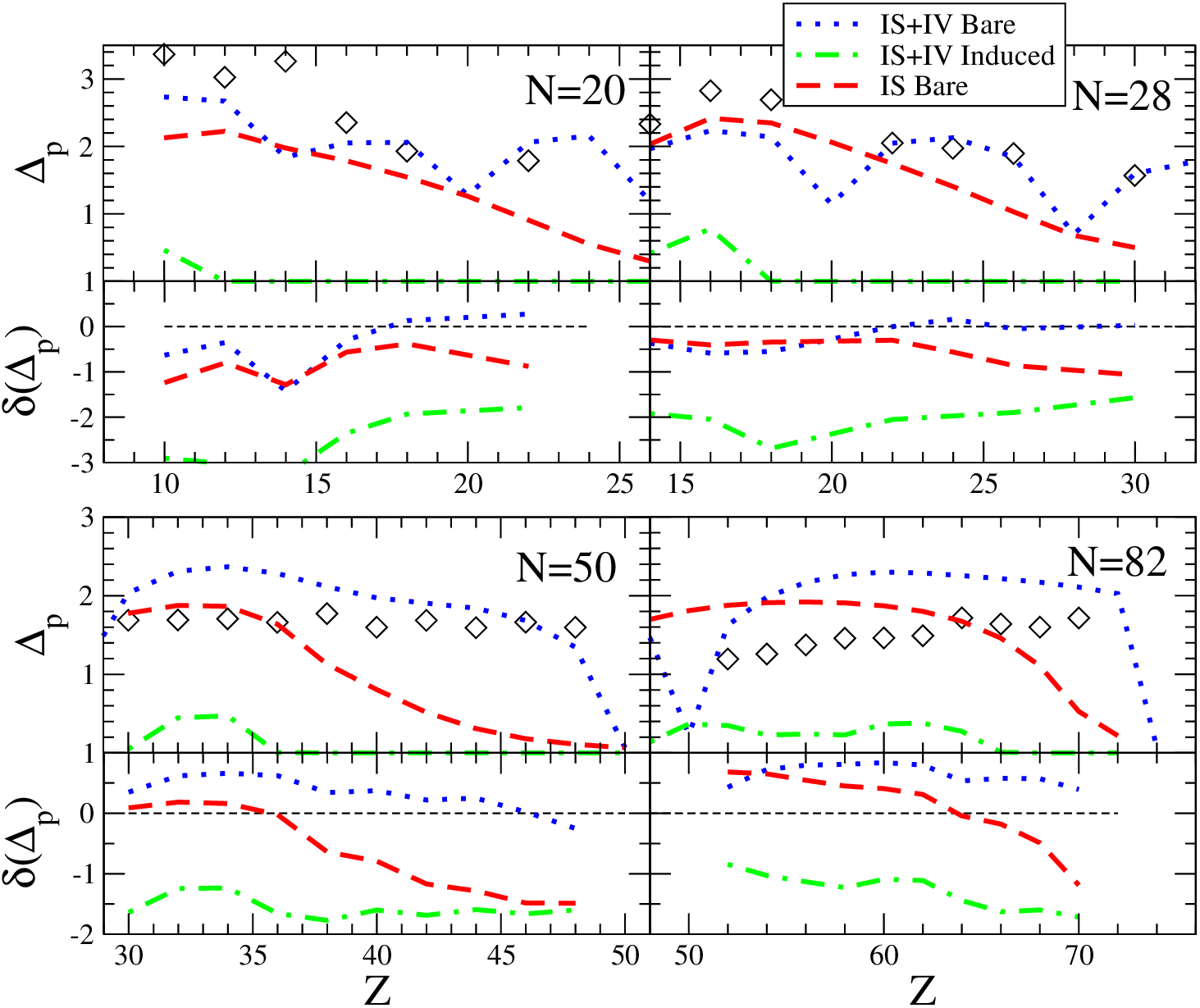}
\end{center}
\caption{(Color online) Comparison of the HFB pairing gaps $\Delta_\mathrm{p}$ calculated with
Eq.~(\ref{eq:delta_mean}) with the OES given by $\Delta^{(3)}$.
The value $\delta(\Delta_\mathrm{p})$ is defined as 
$\delta(\Delta_\mathrm{p})=\Delta_\mathrm{p}(\mathrm{th.})-\Delta_\mathrm{p}(\mathrm{exp.})$.
See the caption of Fig.~\ref{fig10} and the text for details.}
\label{fig13}
\end{figure*}

The proton-proton pairing interaction should also be analyzed in
order to design a global pairing interaction applicable in the whole
nuclear chart, 
It should however be noticed that in our calculations the Coulomb interaction
has not been included in the pairing channel.
The effect of the Coulomb interaction on proton pairing gap has however been 
estimated for instance in Ref.~\cite{hil02} and is expected to decrease the pairing
gap by 100 to 200~keV.
A pertubative estimate of the Coulomb effect on the proton pairing gap has been
evaluated and is expected to be of order of 0.5-1MeV on the pairing gain energy~\cite{ang01}.
This is consistent with the estimation of Ref.~\cite{hil02} for the pairing gap.
Neglecting the Coulomb effect, our calculation is therefore
a semi-quantitative estimate of the proton pairing gap, which
could still be interesting in order to analyze its isotonic dependence.
In Fig.~\ref{fig13} we explore the proton pairing gap in some isotonic chains such as 
e.g. $N$=20, 28, 50 and 82.
The figure shows significant improvement in proton-rich isotones by IS+IV~Bare pairing 
compared with IS~Bare only.
As already observed in the neutron channel, the IS+IV~Induced pairing interaction is not 
strong enough to lead to reasonable proton pairing gaps.

\begin{figure}[tb]
\begin{center}
\includegraphics[scale=0.55]{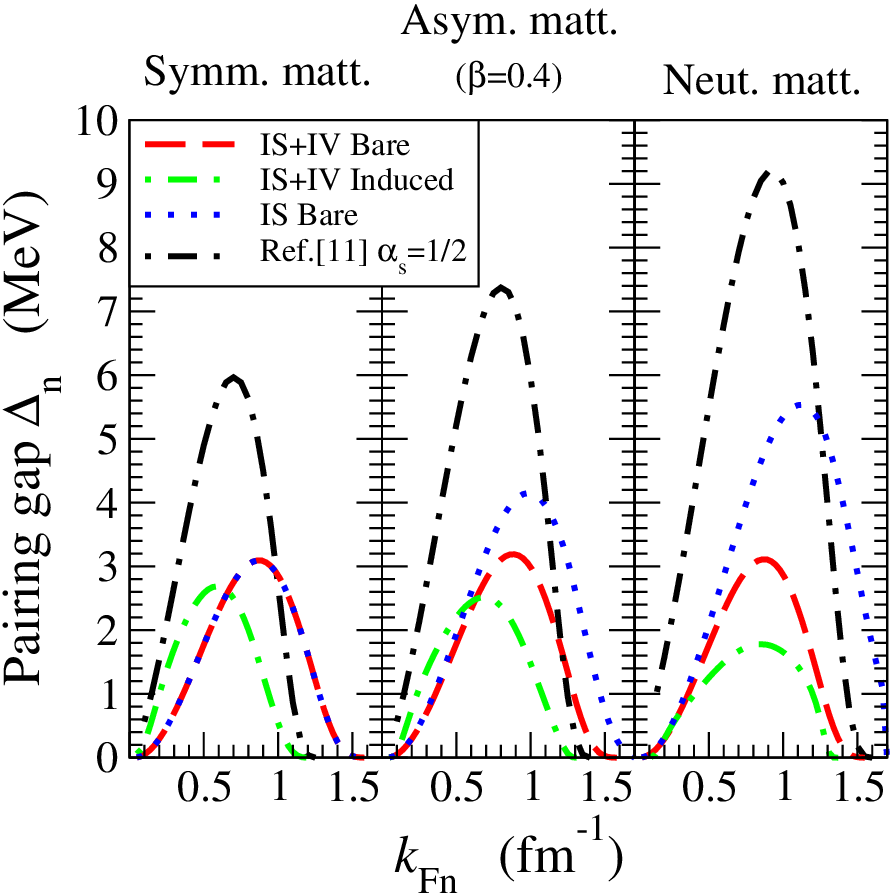}
\end{center}
\caption{(Color online) 
Pairing gaps in uniform matter obtained from the solution of the BCS 
equations with the pairing interactions IS+IV~Bare, IS+IV~Induced, 
IS~Bare and the isoscalar interaction of Ref.~\cite{dob01} with 
$\alpha_\mathrm{s}=1/2$ and $\eta_\mathrm{s}=1$.
Notice that the results of IS+IV~Bare is identical to those of IS~Bare
in symmetric matter. See the text for more details.}
\label{nuc_fig01}
\end{figure}

In order to understand the differences between the interactions IS+IV~Bare and
IS+IV~Induced, we plot in Fig.~\ref{nuc_fig01} the pairing gaps 
in symmetric, asymmetric
(asymmetry parameter $\beta=0.4$) and neutron matter obtained
with these interactions. 
From this figure, it is clear that the pairing gap for symmetric matter 
obtained with the interaction IS+IV~Induced is much smaller than 
that with the interaction IS+IV~Bare for $k_\mathrm{Fn}> 0.7$~fm$^{-1}$,
causing the weak pairing effects in the finite nuclei. 
The medium polarization effects estimated in Ref.~\cite{cao06} shift
the density at the peak position of the pairing gap 
by a factor of $\sim$3 from that of the bare gap, i.e., 
$k_\mathrm{Fn}\sim 0.87$~fm$^{-1}$ ($\rho_\mathrm{n}\sim 0.22\times 10^{-3}$~fm$^{-3}$)
to $k_\mathrm{Fn}\sim 0.60$~fm$^{-1}$ ($\rho_\mathrm{n}\sim 0.73\times 10^{-4}$~fm$^{-3}$).
This change may cause an enhancement of the pairing correlations in very 
low density regime and cause BCS-BEC crossover phenomena~\cite{mar07}.
However, the comparison with the experimental OES shown in 
Fig.~\ref{fig12} clearly indicates that this medium polarization effect 
estimated in infinite matter gives rise to too weak pairing correlations
in finite nuclei. 

Let us now discuss the role of the isovector term.
Since the pairing interaction IS+IV~Bare is in good agreement with
the experimental data as shown in Figs.~\ref{fig10}, \ref{fig11} and
\ref{fig12}, we focus on this pairing interaction. 
To this end, we construct an isoscalar pairing interaction, IS~Bare, 
which is fitted to the bare gap as for the interaction IS+IV~Bare, but
using only the gap in the symmetric nuclear matter.
The parameters for the interaction IS~Bare is listed in Table I (notice 
$\eta_\mathrm{s}=\eta_\mathrm{n}$ and $\alpha_\mathrm{s}=\alpha_\mathrm{n}$). 
The pairing gap in uniform matter obtained with this isoscalar interaction is shown in
Fig.~\ref{nuc_fig01} by the dashed line. 
While in symmetric matter, the interactions IS+IV~Bare and IS~Bare lead to 
identical pairing gaps to each other, the isoscalar interaction IS~Bare produces 
a much larger pairing gap than the IS+IV~Bare interaction, as the asymmetry 
increases.
Moreover, the parameters $\eta_\mathrm{s}$ and $\eta_\mathrm{n}$ in Table~\ref{tab2} 
show that the IS+IV~Bare interaction is of mixed surface and volume 
type in symmetric matter ($\eta_\mathrm{s}$=0.664) as suggested 
in Ref.~\cite{dob02}, and of pure surface type in neutron matter 
($\eta_\mathrm{n}=1.01$), while the IS~Bare pairing interaction is of mixed 
type independently of the asymmetry.
This difference
should manifest itself in the results of finite nuclei.
The binding energy, the two neutrons separation energy, and the average
pairing gap obtained with the interaction IS~Bare are shown by the dashed
line in Figs.~\ref{fig10}, \ref{fig11} and \ref{fig12}, respectively. 
It is clearly seen that while these interactions produce similar
results for $N=Z$ nuclei, the isotopic behavior is somewhat different.
From Fig.\ref{fig12}, it is seen that both the interactions 
IS+IV~Bare and IS~Bare
produce arches of the paring gap in between the neutron magic numbers, but 
the arches induced by the
isoscalar pairing interaction IS~Bare are much larger in amplitude than the one
produced by the interaction IS+IV~Bare.
The difference between the calculated and the experimental pairing
gap, $\delta(\Delta_\mathrm{n})$, estimated with the interaction IS+IV~Bare is indeed
flatter than that with the interaction IS~Bare. 
This behavior suggests clearly the importance of the isovector
component of the pairing interactions as it has already been 
shown for uniform matter in Fig.~\ref{nuc_fig01}. 
We believe that this will bring an important improvement in the
description of pairing in nuclei. 
Further information of the isovector pairing interaction might be 
obtained from experimental study of binding energies of very 
exotic nuclei and excitation spectra of various isotopes.

\begin{figure}[tb]
\begin{center}
\includegraphics[scale=0.51]{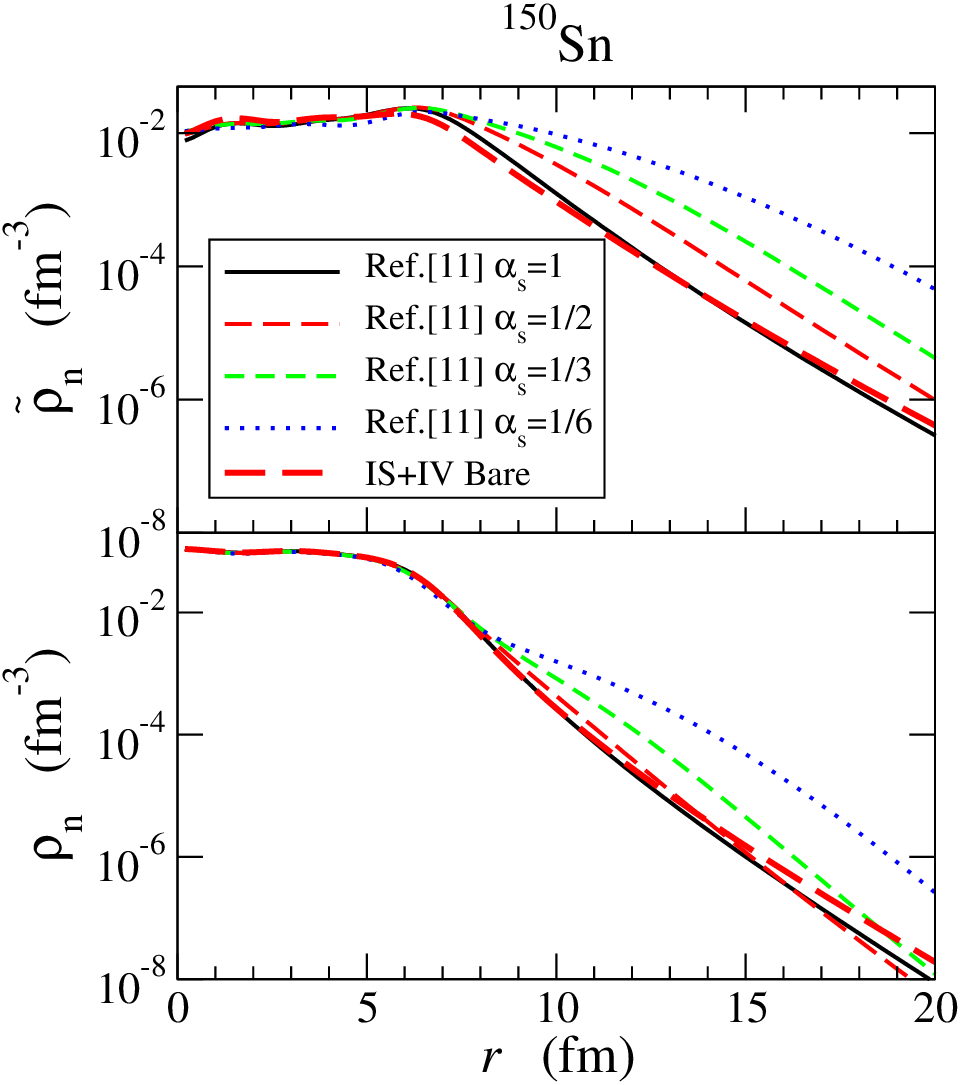}
\end{center}
\caption{(Color online) 
Comparision of 
particle and pairing densities 
for the nucleus $^{150}$Sn obtained with 
several sets of 
interaction in Ref.~\cite{dob01} and with the IS+IV~Bare interaction.
See the text for more details.}
\label{fig08}
\end{figure}

It was pointed out that lower power of the density dependence 
$\alpha_\mathrm{s} < 1/2$ 
with $\eta_\mathrm{s}=1$ in the isoscalar pairing interaction gives rise to anomalous 
behavior in the particle and pairing densities in neutron rich nucleus 
$^{150}$Sn~\cite{dob01}.
In Fig.~\ref{fig08}, the asymptotic behaviour of the particle and pairing 
densities obtained for the set of interactions in Ref.~\cite{dob01} 
are compared with these obtained for the IS+IV~Bare interaction 
in this neutron rich nucleus.
We represent the densities only for the IS+IV~Bare interaction because
in the asymptotic tail, the IS+IV~Bare, IS+IV~Induced 
and IS~Bare are almost undistinguishable.
It is shown that despite the fact that the value of the power of the density
dependence is around $1/2$ for the IS+IV~Bare and IS~Bare interactions,
and less for the IS+IV~Induced interaction,
no anomalous behavior in the densities is observed, contrary to the interactions
studied in Ref.~\cite{dob01}.
We have represented the pairing gaps in symmetric, asymmetric and 
neutron matter for the isoscalar interaction with 
$\alpha_\mathrm{s}=1/2$ and $\eta_\mathrm{s}=1$ in Fig.~\ref{nuc_fig01} (see the
  dot-dashed line).
This interaction induces large values of the pairing gaps at low density from
symmetric to neutron matter.
We have indeed found that the interactions with power of the density dependence
$\alpha_\mathrm{s}$=1/2, 1/3 and 1/6 in Ref.~\cite{dob01} have a peak in the pairing 
gap of absolute value
of about 6~MeV at $k_\mathrm{Fn}\sim 0.7$~fm$^{-1}$ 
($\rho_\mathrm{n}\sim 10^{-2}$~fm$^{-3}$) in symmetric nuclear matter.
The pairing gaps are even increasing when going from symmetric to neutron 
matter, as we already pointed out as a typical behavior for isoscalar pairing 
interactions.
Hence, the anomalous behavior described in Ref.~\cite{dob01} might be
related to an anomalous value of the pairing gap at very low density rather than to
the value of the power of the density dependence of the pairing interaction,
as it was claimed.
In Ref.~\cite{dob01}, the pure surface interactions with $\eta_\mathrm{s}=1$ have 
been adjusted to the value of the pairing gap 
$\Delta_\mathrm{n}=1.25$~MeV in $^{120}$Sn.
From our study, one could conclude that these pure surface interactions 
do not reproduce the pairing gaps in uniform matter obtained from the bare 
microscopic nucleon nucleon interaction.
In order to reproduce them, 
it is indeed necessary to take the parameter $\eta_\mathrm{s}$ as 
adjustable and generate mixed surface and volume pairing interactions, as
it has been done in Refs.~\cite{gar99,mar07,dob02}.

The interaction IS+IV~Bare can be parameterized in a form
\be
\mathrm{g}_\mathrm{n}^3[\rho_\mathrm{n},\rho_\mathrm{p}]
&=&1-\eta_1\left(\frac{\rho_\mathrm{n}}{\rho_0}\right)^{\alpha_1}
-\eta_2\left(\frac{\rho_\mathrm{p}}{\rho_0}\right)^{\alpha_2},
\label{eq:g3n}\\
\mathrm{g}_\mathrm{p}^3[\rho_\mathrm{n},\rho_\mathrm{p}]
&=&1-\eta_1\left(\frac{\rho_\mathrm{p}}{\rho_0}\right)^{\alpha_1}
-\eta_2\left(\frac{\rho_\mathrm{n}}{\rho_0}\right)^{\alpha_2},
\label{eq:g3p}
\ee
with the parameters $\eta_1=1.01$, $\alpha_1=0.525$, $\eta_2=-0.058$,
$\alpha_2=0.559$, and the cutoff energy $E_\mathrm{c}=40$~MeV.
The bare pairing gap could be reproduced by setting $\mathrm{g}_\tau=\mathrm{g}_\tau^3$ in 
Eq.~(\ref{eq:pairing_interaction}).
In the neutron pairing channel, the very weak dependence on the proton density 
is shown from the value of the parameter $\eta_2$ which is close to zero.
With the density-dependent terms~(\ref{eq:g3n}) and (\ref{eq:g3p}), 
we can obtain similar results in finite nuclei to the ones 
obtained with the interaction IS+IV~Bare with the terms (\ref{eq:g1n}) and
(\ref{eq:g1p}).

Let us discuss the qualitative relation between the density-dependent
term $\mathrm{g}_\tau^1$ and $\mathrm{g}_\tau^3$ in the case of the IS+IV~Bare interaction.
Neglecting $\eta_2$ and expressing the variables as
$\rho_\mathrm{n}=(1+\beta)\rho/2$ and $\rho_\mathrm{p}=(1-\beta)\rho/2$
in Eqs.~(\ref{eq:g3n}) and (\ref{eq:g3p}), we obtain to the first order in $\beta$ 
that $\mathrm{g}_\tau^3\approx \mathrm{g}_\tau^1$ if the following relations are respected:
$\eta_\mathrm{s}=\eta_1/2^{\alpha_1}$, $\alpha_\mathrm{s}=\alpha_1$,
$\eta_\mathrm{n}=\eta_\mathrm{s}(1+\alpha_1)$, and $\alpha_\mathrm{n}=\alpha_1$.
These relations provide
a link between the parameters of the density-dependent
terms $\mathrm{g}_\tau^1$ and $\mathrm{g}_\tau^3$.
Moreover, the parameterization (\ref{eq:g3n}) is consistent with
the isospin dependence $\mathrm{f}_\mathrm{s}(\beta)=1-\mathrm{f}_\mathrm{n}(\beta)$ and $\mathrm{f}_\mathrm{n}(\beta)=\beta$ which
is adopted in the present study.


\section{Links between pairing in uniform matter and in nuclei}
\label{sec:lda}


\begin{figure}[b]
\begin{center}
\includegraphics[scale=0.32]{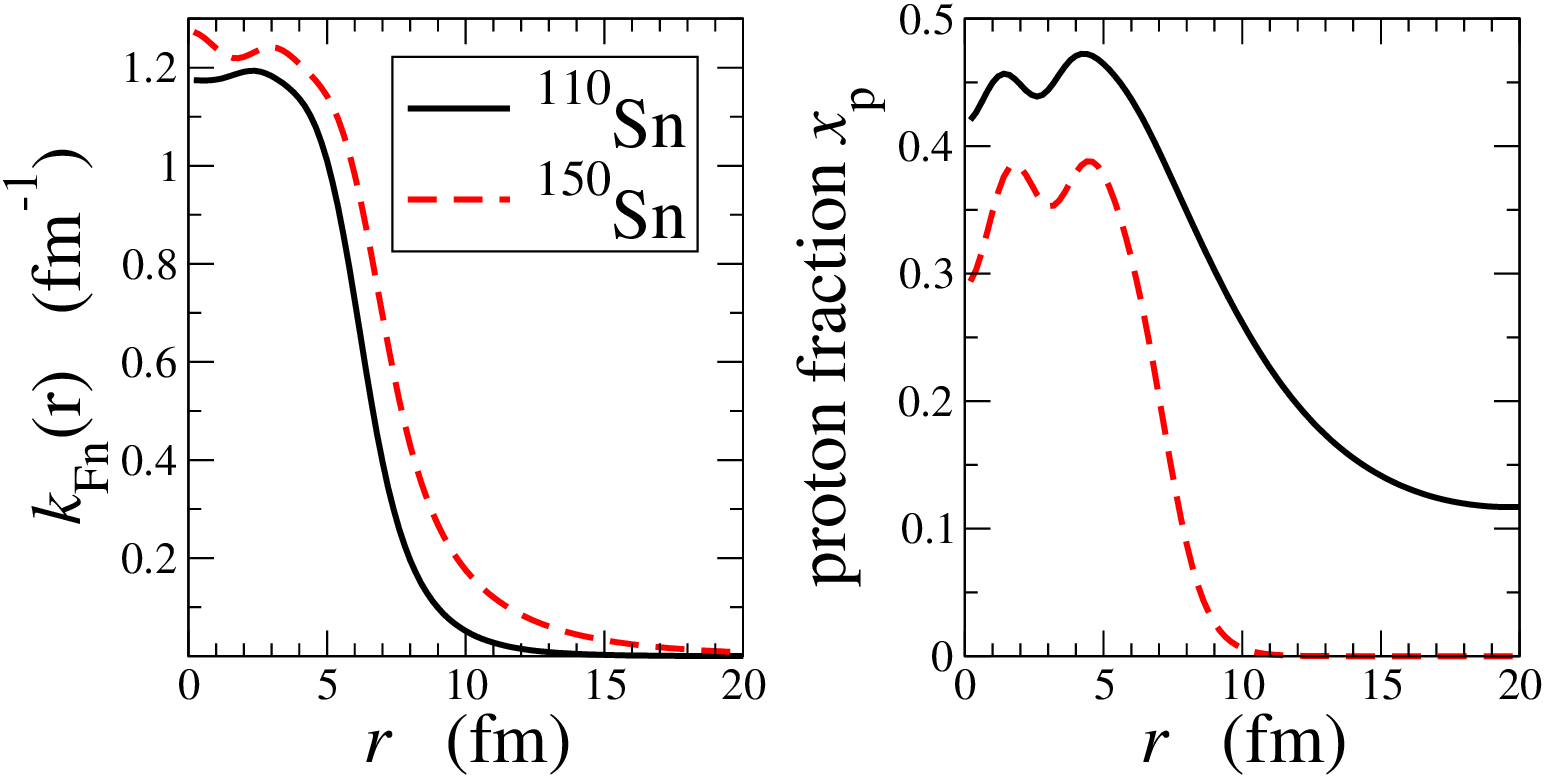}
\end{center}
\caption{(Color online) 
The neutron Fermi momentum and the proton fraction, obtained with the
HFB densities, for two mid-shell Sn nuclei, 
$^{110}$Sn (the solid line) and $^{150}$Sn (the dashed line).}
\label{fig02}
\end{figure}

\begin{figure*}[htb]
\begin{center}
\includegraphics[scale=0.55]{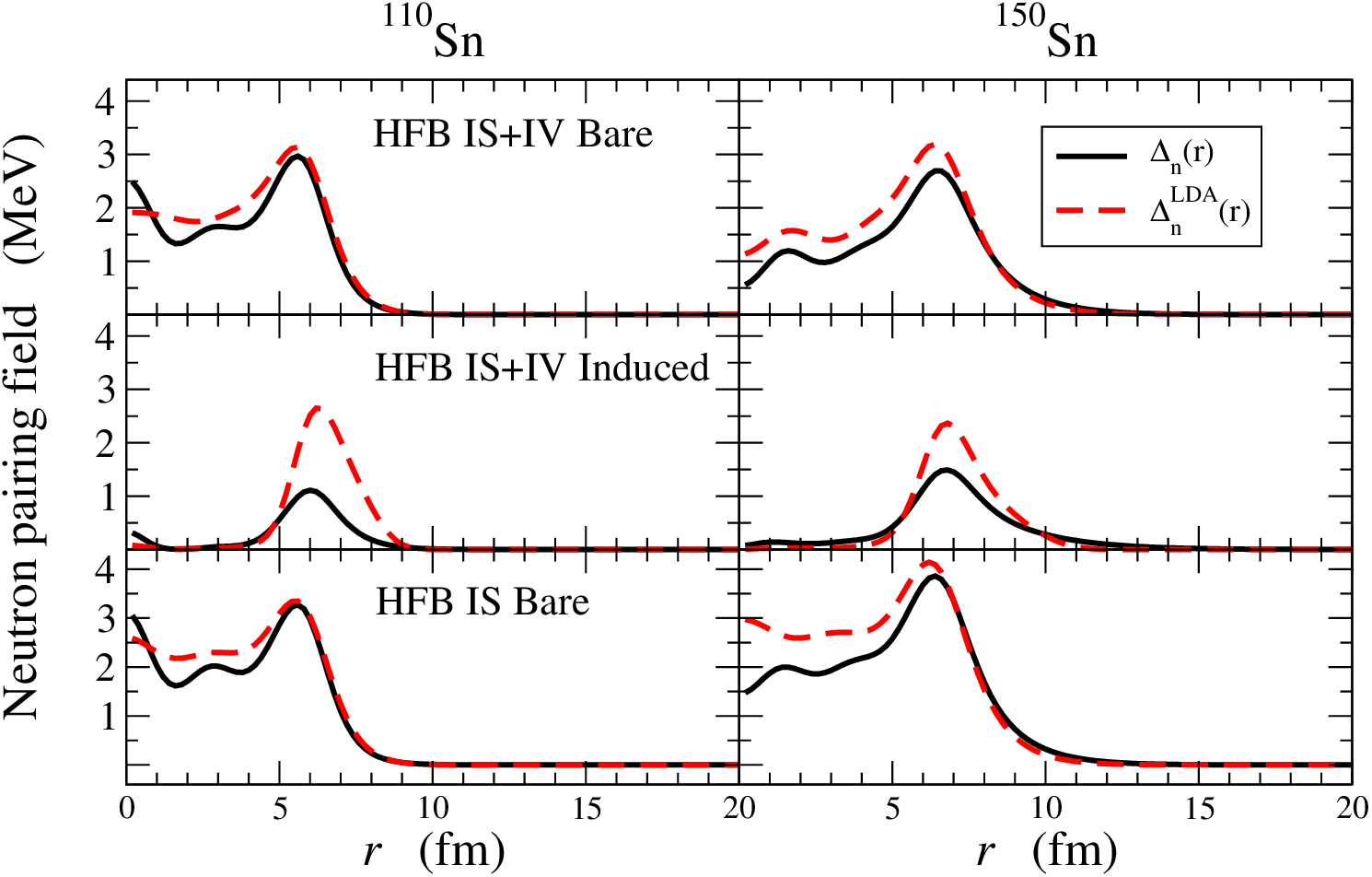}
\end{center}
\caption{(Color online) 
Comparison of the pairing field $\Delta^\mathrm{LDA}_\mathrm{n}(r)$ in the local
density approximation (the dashed line) with that of the HFB
calculations (the solid line) for the 
$^{110}$Sn and $^{150}$Sn nuclei. The pairing fields obtained with 
the different pairing interactions, IS+IV~Bare, IS+IV~Induced and IS~Bare,
are plotted separately.}
\label{fig03}
\end{figure*}

To understand the link between pairing in uniform matter and in
nuclei, let us discuss in this section a local density approximation (LDA) for the pairing
field $\Delta_\mathrm{n}(r)$, defined as 
\be
\Delta_\mathrm{n}^\mathrm{LDA}(r)\equiv\Delta_\mathrm{n}^\mathrm{unif}
\left(k_\mathrm{Fn}(r), x_\mathrm{p}(r)\right) \; ,
\ee
where $\Delta_\mathrm{n}^\mathrm{unif}(k_\mathrm{Fn}, x_\mathrm{p})$ 
is the pairing gap in uniform matter
calculated for a given Fermi momentum $k_\mathrm{Fn}$ and proton fraction $x_\mathrm{p}$.
The LDA consists in replacing these variables by the local ones defined 
in finite nuclei. 
The local Fermi momentum $k_\mathrm{Fn}(r)$ and the local proton fraction 
$x_\mathrm{p}(r)$ are thus defined as,
\be
k_\mathrm{Fn}(r)&=&\left(3\pi^2\rho_\mathrm{n}(r)\right)^{1/3} \; , \\
x_\mathrm{p}(r)&=&\rho_\mathrm{p}(r)/\left(\rho_\mathrm{n}(r)+\rho_\mathrm{p}(r)\right) \; .
\ee
The neutron and proton densities, $\rho_\mathrm{n}(r)$ and
$\rho_\mathrm{p}(r)$, are given by the HFB calculation in finite nuclei. 
We represent $k_\mathrm{Fn}(r)$ and $x_\mathrm{p}(r)$ 
for two mid-shell nuclei, $^{110}$Sn and $^{150}$Sn in Fig.~\ref{fig02}. 
At the surface of the nuclei, the proton fraction is decreasing faster in 
$^{150}$Sn than in $^{110}$Sn and the local Fermi momentum
$k_\mathrm{Fn}(r)$ is slightly larger in $^{150}$Sn than in $^{110}$Sn.
Then, if pairing correlations are important at the surface,
the pairing fields $\Delta_\mathrm{n}^\mathrm{LDA}(r)$ in the LDA should depend 
on the isospin properties of the pairing interaction.
For these nuclei, the pairing fields in the LDA are shown 
in Fig.~\ref{fig03} for each of the pairing interactions IS+IV~Bare,
IS+IV~Induced and IS~Bare. 
To this end, we have calculated the solution of the BCS equations in 
asymmetric matter~\cite{mar07} and used it as the pairing gap in the 
uniform matter $\Delta_\mathrm{n}^\mathrm{unif}(k_\mathrm{Fn},x_\mathrm{p})$, that 
is the same as the ones represented in
Fig.~\ref{nuc_fig01}. 
For a comparison, the pairing fields $\Delta_\mathrm{n}(r)$
obtained with the HFB calculations are also plotted in 
Fig.~\ref{fig03}. 
It is clear from the figure that the pairing interactions IS+IV~Bare and
IS~Bare have a mixed character of surface and volume types, while the
pairing field obtained with the interaction IS+IV~Induced is strongly surface peaked.
It is surprising that the LDA provides not only qualitative 
but also quantitative description of the pairing field in finite nuclei.
Nevertheless, finite size effects which are neglected in the
LDA are not negligible, and the LDA overestimates the pairing 
field by about 10-20\% for the pairing interactions of the mixed volume 
and surface type correlations, like the IS+IV~Bare and IS~Bare ones, 
and by 50\% for the pure surface type pairing correlations, like the
IS+IV~Induced interaction.
From the pairing field, one could deduce the mean pairing gap
according to Eq.~(\ref{eq:delta_mean}).  
We show in Table~\ref{tab3} those mean pairing gaps obtained for 
$^{150}$Sn for the set of pairing interactions IS+IV~Bare, IS+IV~Induced,
and IS~Bare.
In addition, we also calculate the pairing gap with another
expression, 
\be
\tilde{\Delta}_\mathrm{n}\equiv\frac{1}{\tilde{N}}\int d^3r \tilde{\rho}_\mathrm{n}(r)
\Delta_\mathrm{n}(r) \; ,  
\label{eq:adelta_mean} 
\ee
where $\tilde{N}\equiv\int d^3r\tilde{\rho}_\mathrm{n}(r)$ is the average number of
neutrons participating to the pairing field.
As expected, the average pairing gap is overestimated in the LDA
approximation. 
For the interaction IS+IV~Induced, the LDA even predicts a pairing gap smaller
than the ``experimental'' one.
It should also be remarked from the Table~\ref{tab3} that the average pairing
gaps $\Delta_\mathrm{n}$ and $\tilde{\Delta}_\mathrm{n}$ are very similar for the surface and 
volume mixed-type pairing interactions IS+IV~Bare and IS~Bare, while there are
important differences for the surface peaked interaction IS+IV~Induced.

\begin{table}[tb]
\begin{center}
\setlength{\tabcolsep}{.06in}
\renewcommand{\arraystretch}{1.5}
\begin{tabular}{clcccc}
\toprule 
nuclei & pairing       & \multicolumn{2}{c}{HFB} & \multicolumn{2}{c}{LDA} \\ 
       & interactions  & $\Delta_\mathrm{n}$ & $\tilde{\Delta}_\mathrm{n}$ &
$\Delta_\mathrm{n}^\mathrm{LDA}$ & $\tilde{\Delta}_\mathrm{n}^\mathrm{LDA}$ \\[-0.1cm]
 & & (MeV) & (MeV) & (MeV) & (MeV) \\
\colrule
$^{110}$Sn & IS+IV~Bare & 2.04 & 2.07 & 2.33 & 2.32 \\
&IS+IV~Induced & 0.39 & 0.57 & 0.68 & 1.19 \\
&IS~Bare & 2.32 & 2.32 & 2.60 & 2.53 \\[0.1cm]
$^{150}$Sn & IS+IV~Bare & 1.72 & 1.71 & 2.16 & 2.04 \\
& IS+IV~Induced & 0.61 & 0.77 & 0.76 & 1.06 \\
& IS~Bare & 2.67 & 2.53 & 3.13 & 2.79 \\
\botrule 
\end{tabular}
\end{center}
\caption{The mean pairing gap $\Delta_\mathrm{n}$ and $\tilde{\Delta}_\mathrm{n}$
  for $^{110}$Sn and $^{150}$Sn calculated with 
  Eqs.~(\ref{eq:delta_mean}) and (\ref{eq:adelta_mean}), respectively.
These are obtained by using either   
  either the HFB pairing field $\Delta_\mathrm{n}(r)$ or the 
LDA pairing filed $\Delta_\mathrm{n}^\mathrm{LDA}(r)$ for the three density-dependent 
pairing interactions, IS+IV~Bare, IS+IV~Induced and IS~Bare. }
\label{tab3}
\end{table}%

Notice that 
these LDA results are model dependent in a sense that they rely on a
model for the neutron and proton density profile. However, except for this
aspect, the local density approximation is qualitatively justified and
the variation of the densities due to the pairing correlations are 
small. 
Presumably, one can consider that the LDA provides a reliable 
tool for a qualitative understanding of the pairing correlation 
in finite nuclei. 


\section{Conclusions}
\label{sec:conc}


We have performed HFB calculations for semi-magic Ca, Ni, Sn, and Pb isotopes 
and $N$=20, 28, 50 and 82 isotones
using the density-dependent pairing interactions~\cite{mar07}
deduced from microscopic nucleon-nucleon interaction~\cite{cao06}.
Three interactions have been employed, namely, isospin dependent 
interactions adjusted to the pairing gaps both in 
symmetric and neutron matter obtained from the bare nucleon interaction
(IS+IV~Bare), or to those modified by medium polarization 
effects (IS+IV~Induced),
and an isoscalar interaction adjusted only in symmetric matter 
(IS~Bare). 
We have compared the results of these pairing interactions with 
the experimental data for binding energies, 
two neutrons separation energies, and odd-even mass staggering (OES).

We have found that the two pairing interactions IS+IV~Bare and IS+IV~Induced
lead to different results in finite nuclei. 
The comparison with the experimental OES suggests that the experimental data 
favor the interaction IS+IV~Bare, which reproduces the bare pairing gap in both
symmetric and neutron matter.
These results indicate that the medium polarization effects estimated in 
infinite matter provides weaker pairing correlations than
observed in finite nuclei.
The discrepancies concerning the role of the phonon coupling
between the calculations presented in
Ref.~\cite{bar99,gio02,bar05} for finite nuclei and the calculations
for uniform matter in Ref.~\cite{cao06}, therefore, still remain
an open question. 

An interesting result shown in this paper is that the pairing 
interaction IS+IV~Bare leads to good agreements with the 
experimental masses for light, intermediate and heavy nuclei 
without any tuning in different isotopes.
This suggests that an inclusion of the isovector term in the effective
pairing interaction helps in designing a global interaction applicable 
in the whole nuclear chart, taking advantage of the simplicity of the 
contact pairing interaction.
It should however be noticed that in the proton pairing channel, the 
Coulomb interaction has not yet been included in our calculations.
This should be done in futur investigations.

We have shown that the anomalous behavior of
particle and pairing 
densities obtained in Ref.~\cite{dob01} for isosclar pairing 
interactions of surface type with the power of the density 
dependence $\alpha_\mathrm{s} < 1/2$ is related to the large pairing gaps 
generated by these interactions at low density.
The volume and surface mixed-type interaction adopted in the
present study does not show this anomaly despite that the parameter
$\alpha_\mathrm{s}$ is close to $1/2$.

Finally, we have discussed the local density approximation (LDA) 
for the pairing field, and have shown that it leads to a nice qualitative 
description of the pairing correlations in finite nuclei. 
The comparison of the pairing field obtained from the HFB 
calculation with the one extracted using the LDA suggests that there
is a possibility to map from the pairing in uniform matter to that
in finite nuclei.

\acknowledgments
One of us (J.M.) would like
to thank M.~Grasso for her help in the initial 
stage of this study.
We also thank M.~Matsuo, K.~Matsuyanagi and W.~Nazarewicz 
for usefull discussions.
This work was supported by the Japanese
Ministry of Education, Culture, Sports, Science and Technology
by Grant-in-Aid for Scientific Research under
the Program No. 19740115.


\end{document}